\documentclass[
amsmath, 
amssymb, 
aps, 
pra,
twocolumn,
floatfix,
nofootinbib
]{revtex4}
\usepackage[utf8]{inputenc}
\usepackage{titlesec}
\usepackage[colorlinks]{hyperref}
\usepackage{amsmath,amssymb}
\usepackage{pifont}
\usepackage{bbold}
\usepackage{tabularx}
\usepackage{float}
\usepackage{dcolumn}
\usepackage{bm}
\usepackage{dsfont}
\usepackage{url}
\usepackage{mathtools}
\usepackage{upgreek}
\usepackage{booktabs}
\usepackage[title]{appendix}

\usepackage{amsfonts}
\usepackage{braket}
\usepackage{cancel}
\usepackage{bbold}
\usepackage{makecell}
\usepackage[normalem]{ulem}
\usepackage{xcolor}
\usepackage{comment}

\usepackage{afterpage}
\newcommand{\xmark}{\ding{55}}

\begin{document}
\title{Boosting Quantum Key Distribution via the End-To-End Loss Control}

\author{A.\,D.\,Kodukhov$^{1,\dag}$}

\author{V.\,A.\,Pastushenko$^{1,\dag}$}

\author{N.\,S.\,Kirsanov$^1$}

\author{D.\,A.\,Kronberg$^1$}

\author{M.\,Pflitsch$^1$}

\author{V.\,M.\,Vinokur$^{1,*}$}

\address{\mbox{$^1\,$Terra Quantum AG, Kornhausstrasse 25, 9000 St.\,Gallen, Switzerland}
\\$^\dag\,$these authors contributed equally to this work
\\$*\,$vv@terraquantum.swiss
}

\begin{abstract}
With the rise of quantum technologies, data security increasingly relies on quantum cryptography and its most notable application, quantum key distribution (QKD). 
Yet, current technological limitations, in particular, the unavailability of quantum repeaters, cause
relatively low key distribution rates in practical QKD implementations. 
Here, we demonstrate a remarkable improvement in the QKD performance using end-to-end line tomography for the wide class of relevant protocols. 
Our approach is based on the real-time detection of interventions in the transmission channel, enabling
an adaptive response that modifies the QKD setup and post-processing parameters, leading, thereby, to a substantial increase in the key distribution rates. 
Our findings provide everlastingly secure efficient quantum cryptography deployment potentially overcoming the repeaterless rate-distance limit.
\end{abstract}

%\keyword{quantum key distribution, optical fiber, Rayleigh scattering, loss control, transmittometry, optical time-domain reflectometry, line tomography, privacy amplification.} 

\maketitle

\section{Introduction}
As we step towards the quantum computing era, quantum cryptography is emerging as the primary solution for ensuring data security. 
A cornerstone of quantum cryptography is quantum key distribution (QKD). 
This technique harnesses unique quantum properties, such as the no-cloning property of quantum states and quantum entanglement, to securely share encryption keys between two or more parties. 
Paired with the symmetric classical cryptography routines, the QKD becomes a silver bullet against the security threats posed by quantum computers.
Yet, deploying the QKD in the real world meets strong challenges. 
While the QKD protocols, such as BB84, boast theoretical security, their practical implementations
face a wide range of challenges.
in particular, the~practical realization of quantum transmission channels is marred by substantial signal losses, and~it is assumed that an eavesdropper, Eve, can seize and manipulate these losses to her advantage.
Here, we show how physical control over the channel losses can ward off the known attacks
targeted at exploiting those losses against the existing prepare-and-measure QKD protocols, in~particular, Decoy-State BB84\,\cite{BB84_orig,hwang2003quantum,Decoy1,Decoy2, trushechkin2021_decoy} and COW\,\cite{COWstucki2005fast, COWstucki2009high, COWkorzh2015provably}.
Furthermore, we find how this physical loss control can be used to overcome the fundamental PLOB (Pirandola-Laurenza-Ottaviani-Banchi) bound\,\cite{PLOB}. 
This boundary predicts an exponential decrease in key rates in correspondence with increasing communication distance, posing a significant limitation to long-distance secure quantum communications.  

Our approach shifts the conventional quantum cryptography paradigm, which assumes that an eavesdropper, Eve, can capture and exploit all losses occurring in the transmission channel. 
We state that during the transmission along the optical fiber, most of the signal losses occur due to scattering on the fiber density fluctuations in the channel and that it is practically impossible to collect these scattered losses. 
Thus, by~complementing quantum mechanical restriction imposed on a potential eavesdropper with realistic restrictions stemming from the development of technology, we narrow the class of attacks that have to be considered down to attacks utilizing deliberate local interventions.
Moreover, relying on the end-to-end loss control method proposed in\,\cite{forty}, legitimate users can detect these local intrusions.
Remarkably, we find that our approach enables legitimate users to employ higher signal intensities and radically improves key distribution~rates.

We demonstrate the high quantum cryptography potential that can be realized utilizing the physical end-to-end control without making strong assumptions about the eavesdropper's capabilities. 
These 
assumptions 
will be explored in depth in \mbox{Sections\,\ref{sec:natural} and \ref{sec:control}.} 
Section\,\ref{sec:an_sc} describes the scheme utilized in Sections\,\ref{sec:BB84} and \ref{sec:COW} for analyzing the influence of the proposed method on BB84 and COW QKD protocols, respectively.
\mbox{Section\,\ref{sec:comparison}} provides the comparison of the proposed loss control technique with the conventional decoy-state approach, Section\,\ref{sec:beyond} touches upon the limitations of our approach and, finally, Section\,\ref{sec:discussion} presents our~conclusions.

\section{Eavesdropping of natural losses}\label{sec:natural}

Most of the QKD realizations leverage telecom equipment and employ optical fiber as a transmission channel.
The losses in the fiber channel stem predominantly from the Rayleigh scattering, caused by homogeneously distributed quenched disorder.
Building on Quantum Thermodynamics considerations, see Ref.\,\cite{forty}, we argue that these\,\textit{natural} losses cannot be effectively harvested by Eve.
We carry out an illustrative analysis, the~results of which show that efficient eavesdropping of natural losses would require the length of the collection apparatus that is impossible to practically realize at the present level of~technology.

In the most QKD protocols, the classical information is encoded via the parameters of the coherent states. 
We thus set that the bits 0 and 1 are encoded by the coherent states $\ket{\gamma_{0}}$ and $\ket{\gamma_{1}}$, respectively.
Let us consider a fiber channel segment of the length $l$.
In the absence of local leakages, the~fraction of the signal scattered at this segment is given by
\begin{equation}
    r_l = 1-10^{-\xi l},   
\end{equation}
where $\xi=0.02\,\text{km}^{-1}$ is the attenuation constant typical for fiber.
Then, in~order to obtain information,
Eve 
has to be able to distinguish
between the equiprobable effective lost components $\ket{\sqrt{r_l}\gamma_{0}}$ and $\ket{\sqrt{r_l}\gamma_{1}}$.
The maximum amount of information $I_l$ that can be extracted from these states is upper-bounded by the fundamental Holevo quantity  $\chi$\,\cite{Holevo}, which in this case can be written as
\begin{equation}
    I_l \leq \chi= h_2\left( \frac{1-|\braket{\sqrt{r_l}\gamma_0|\sqrt{r_l}\gamma_1}|}{2} \right),
    \label{PureStateBound}
\end{equation}
where $h_2(x)=-x\log_2x-(1-x)\log_2(1-x)$ is binary entropy.

\begin{figure}[t]
\noindent\centering{
\includegraphics[width=0.99\columnwidth]{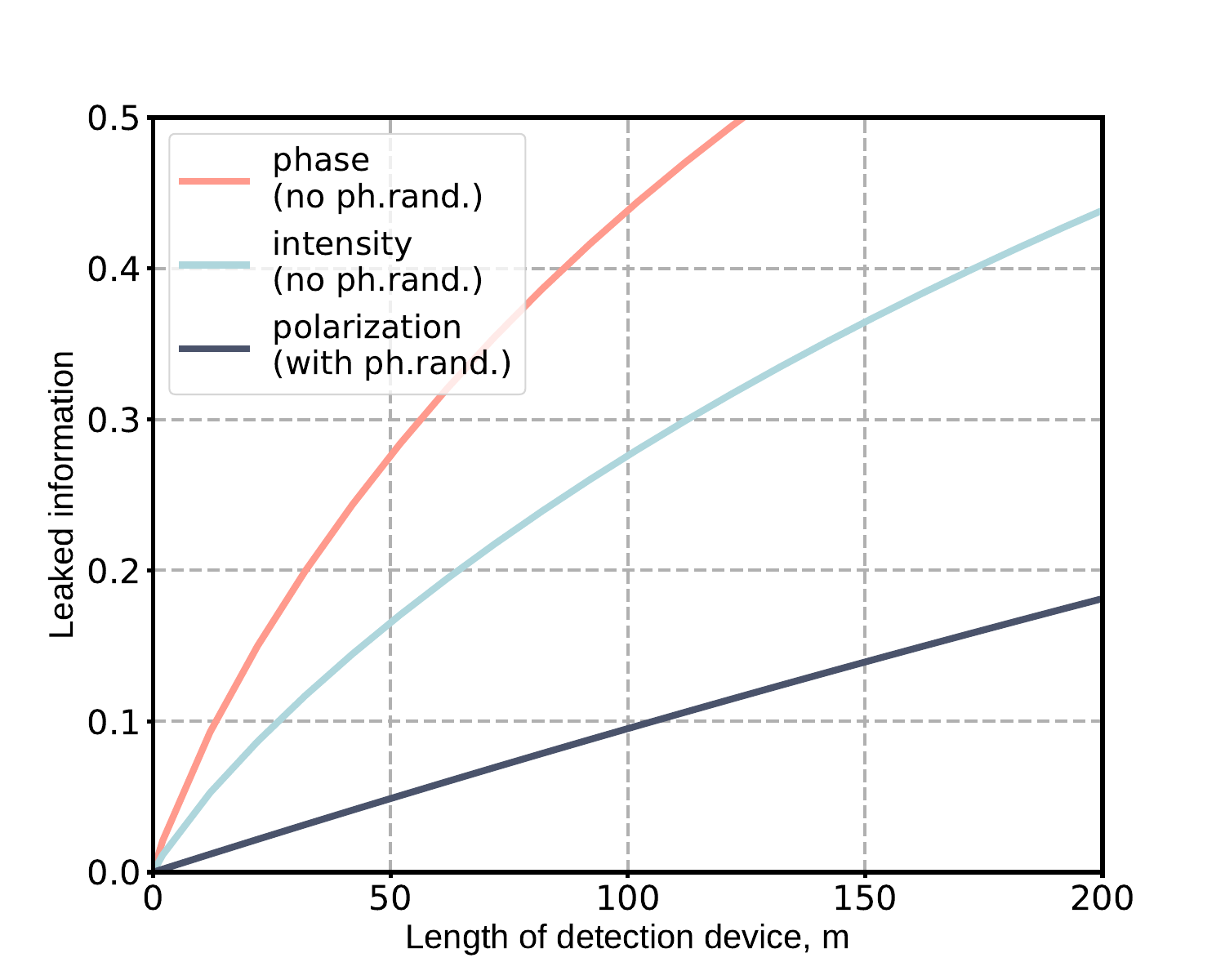}}
\caption{
\textbf{Information leaking from natural losses.}
The amount of information per pulse that Eve can obtain from natural losses as a function of the overall length of the detection device.
The coral line depicts the information estimated according to Equation\,(\ref{PureStateBound}) for  a protocol utilizing the following encoding scheme: ``0''$\rightarrow\ket{\gamma}\ket{\gamma}$, ``1''$\rightarrow\ket{\gamma}\ket{-\gamma}$, which resembles DPS protocol.
The cyan line corresponds to the case of COW protocol: ``0''$\rightarrow\ket{0}\ket{\gamma}$, ``1''$\rightarrow\ket{\gamma}\ket{0}$ --- the estimation was also conducted according to Equation\,(\ref{PureStateBound}).
The dark blue line depicts the estimation Equation\,(\ref{PRBound}) built for the encoding into polarizations of phase-randomized coherent states of amplitude $\gamma$, i.e.,~for BB84 protocol.
For all encoding methods, the~average number of photons $|\gamma|^2=100$ that appeared to be optimal for bit-encoding states in the context of our approach (see Figures\,\ref{BB84number} and \ref{COWnumber}).}
\label{NaturalLossesDetector}
\end{figure}

When considering protocols that utilize phase-randomized coherent states, such as practical realizations of BB84, where information is carried by polarization, we assume that Eve acquires a bit each time when a non-zero count of photons is retrieved from the scattered signal.
Thus, the average information associated with the bit is defined by the Poisson~statistics  

\begin{equation}
    I_l \leq 1-e^{-r_l|\gamma|^2},
    \label{PRBound}
\end{equation}
where $|\gamma|^2$ is the average photon number in each signal pulse regardless of the encoded bit value.

Figure\,\ref{NaturalLossesDetector} depicts the information leakage, attributable to Rayleigh scattering, as a function of $l$.
Three curves correspond to distinct types of protocols, differing by the physical parameters selected for information encoding, the intensity, phase, and polarization.
We find that to gain a considerable informational advantage over the legitimate users, Eve would need to cover a line segment exceeding a hundred meters.
This condition makes an undetectable attack utilizing the natural losses unfeasible.

Notably, in~the case of the bit-encoding states with equal energies, like polarization or phase encoding methods, the~provided upper bounds are significantly higher than the amount of information that can be practically extracted.
In principle, a~potential eavesdropper may measure the losses from each scattering center individually.
However, the~resulting precision will be completely obscured by the quantum noise, and~the obtained information will be much smaller than information estimations of Equations\,(\ref{PureStateBound}) and (\ref{PRBound}).
Thus, a~collective measurement unifying, for~efficacy, all occurring losses into one narrow wave packet with the amplitude $\sqrt{r_l}\gamma_{0,1}$ is necessary for approaching the upper bounds.
This problem is comparable, in~its complexity, with~reversing the evolution of a scattered wavefront\,\cite{lesovik2016h,lesovik2019arrow,kirsanov2018entropy,kirsanov2018h}.

\section{Line tomography}\label{sec:control}
The line tomography as a constituent part of the key distribution process was originally proposed and thoroughly discussed in Ref.\,\cite{forty}. 
In this section, we briefly outline its basis and discuss its inalienable components.
In our approach, losses other than natural are monitored through line tomography.
By accurately quantifying the exploitable leakages, users precisely identify the amount of information potentially intercepted by Eve. 
This knowledge enables them to execute the most efficient privacy amplification procedure, thus, enhancing key distribution rates. 
Furthermore, it allows Alice and Bob to appropriately modify the parameters of the bit-encoding quantum states, making them less discernible to Eve, which, as~we will show later, boosts the key rates even~further.

Line tomography involves two distinct  procedures: Optical Time-Domain Reflectometry (OTDR) and transmittometry, both of which contribute to a comprehensive knowledge of the losses in the line.
Each component utilizes high-intensity test pulses which are dispatched at high frequencies, running concurrently with the bit-encoding~states.

\begin{figure}[t]
\noindent\centering{
\includegraphics[width=0.99\columnwidth]{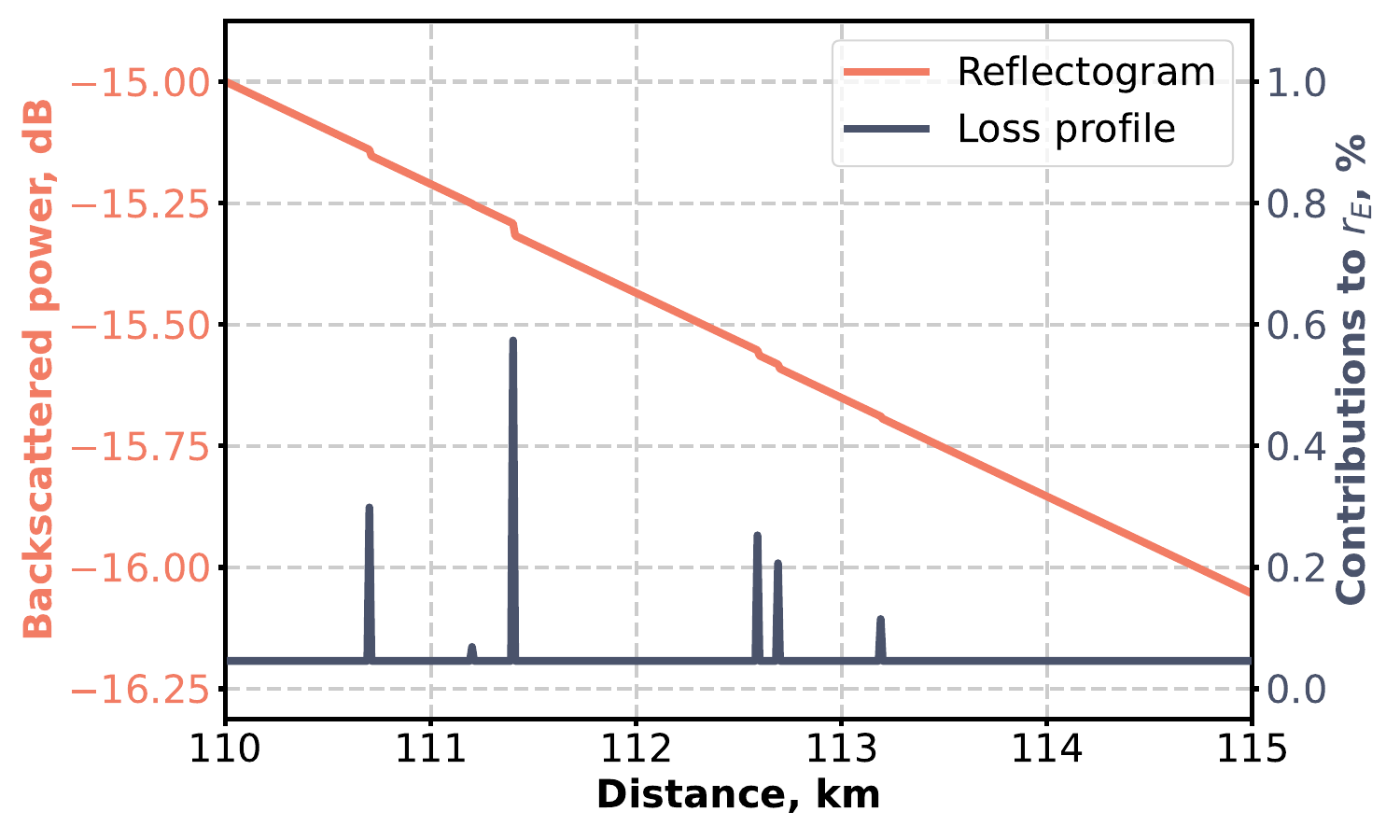}}
\caption{
\textbf{Exemplary reflectogram and line tomogram.}
The loss profile, which displays the magnitude of the $i$-th local leakage and its position, is derived from the reflectogram.}
\label{LossProfile}
\end{figure}

The OTDR is based on recording backscattered optical radiation from test pulses that travel through the fiber. 
The distance to a particular scattering point is calculated by measuring the time delay of its arrival, while the intensity of the backscattered signal provides information about the magnitude of any detected leakage.
Figure\,\ref{LossProfile} sketches an exemplary reflectogram (upper trace) and the corresponding loss tomogram (lower trace). 
The tomogram is derived from the reflectogram by fitting it with a combination of a linear decline function, representing natural losses (keep in mind that the reflectogram is a semi-log plot), and~weighted step-like functions. 
The tomogram maps the discrete derivative values of the step-like functions to the respective positions, thus pinpointing the local leakages.
To achieve the loss detection accuracy of 0.5\% and better test pulses must comprise about $10^{11}$ photons, while the time duration of a pulse is about 1\,$\upmu$s
 and wavelength $\lambda$ is 1530\,nm.

At the same time, transmittometry detects the transmitted components of the test pulses, enabling a cross-comparison of input and output intensities between the users. 
Although this method does not allow users to identify the exact locations of local leakages, it does provide the total leakage value $r_\text{E}$, calculated using the a priori known baseline of natural losses.
Transmittometry can be further enhanced by modulating test pulses at high frequencies, similar to what is used in the lock-in technique. 
The test pulses’ modulation is primarily needed to suppress the $1/f$ low-frequency noise of the laser emission. Analyzing the peaks of input and output spectral power corresponding to the modulation frequency ($\sim$MGz) allows users to calculate the lost proportion of the light. The~$1/f$ noise is effectively suppressed, which enhances the accuracy of transmittometry.
The fiber structure and hence line tomogram are physically unclonable functions\,\cite{PUF_opt_fib}.
Thus, global actions by Eve, such as substituting the line with a lossless channel or completely blocking certain signals, can be readily detected due to their significant impact on the line tomogram. We discuss this issue in detail in\,\cite{forty, PUF_opt_fib}. 
Consequently, these drastic Eve's actions would not provide an efficient method to interfere, since in case of these actions, the~key generation will be immediately terminated by the legitimate users.
Therefore, Eve's interventions are limited to creating minor local leakages which, however, are accurately measured by users and, correspondingly, are taken into account.
Note that Eve is not able to selectively skip the high-intensity test pulses while being able to seize the low-intensity signal pulses, since physical detection of intensity requires her direct installation of splitters into the line.
This would induce significant losses and significantly affect the reflectorgram of the line for a tangible period of time detectable by transmitometry.
Moreover, measuring signal intensities causes time delays in the signal transmission from Alice to Bob which is also~noticeable.

The primary constraint of the approach pertains to the accuracy of the loss control. 
Reflectometers are characterized by their resolution, which enables the localization of line discontinuities and the documentation of silica structure. 
If the resolution is not sufficient, the~reflectogram, serving as an unclonable physical fingerprint of the line, would not enable registering some of the very localized anomalies in the line. 
Accumulating sufficient data to construct a meaningful reflectogram can also be time-intensive. 
Transmittometry provides an immediate update of the total effective loss magnitude, even before the reflectogram is fully constructed. 
Yet, its precision is restricted by line noises that need to be mitigated. 
Another limitation emerges when local line losses exceed a certain threshold at any point. 
In such cases, the~transmission must be paused and the line inspected. This level of loss can indicate the potential installation of an interception equipment by~Eve. 

It is important to note that before starting the key generation process, Alice and Bob perform line tomography to determine the natural losses in the channel.
Only at this preliminary step, the~legitimate users must be certain that Eve does not introduce additional losses in the line.
Eve may also exploit some of the predetermined leakages in the line, e.g.,~losses at connectors or fiber bends.
For a particular line, such leakages have to be taken into account and added to the total leakage $r_\text{E}$.

\section{Analysis scheme}\label{sec:an_sc}
Line tomography can be applied to a wide range of the QKD protocols and
enable legitimate users to optimize the parameters of the bit-encoding states and post-processing and, thus, to~significantly enhance the key rate, all while maintaining the same security threshold.
Modified intensities of the bit-encoding pulses depend on the loss detection accuracy and may reach $10^2$ photons per pulse (see Figures\,\ref{BB84number} and \ref{COWnumber} for details).
However, the~specifics of each protocol do affect how the improved key rates are calculated.
Depending on whether the protocol incorporates randomization of the bit-encoding states' phases\,\cite{PhaseRand1, PNS_Lutkenhaus, PhaseRand3, PhaseRand4} or not, the~prepare-and-measure QKD protocols can be divided into two~categories.

The first category includes protocols that employ quantum systems having a finite Hilbert space, specifically, qubits.
In the ideal scenario, these protocols would use single photons for encoding.
Despite the latest achievements in single photon sources for QKD, e.g.,~based on quantum dots\,\cite{single_photon_source}, the~weak phase-randomized coherent states are often utilized in the practical realization of these protocols, since they are more available and easier to exploit.
The well-known examples of protocols of this group include the Decoy-State BB84\,\cite{BB84_orig,hwang2003quantum,Decoy1,Decoy2,trushechkin2021_decoy,exp_BB84_1,exp_BB84_2}, Six-State\,\cite{SixState}, T-12\,\cite{T12,T12_experiment}, and~SARG04\,\cite{SARG04}.
One of the most powerful attacks on these protocols is the photon number splitting (PNS) attack\,\cite{PNS_attack_2000, Acin_Gisin_Scarani, SARG04, trushechkin2021_decoy} which allows Eve to collect surplus photons from the multiphoton pulses and forward the rest photons to Bob via a lossless channel.
This attack provides an upper bound for the secret key rate, which, remarkably, together with the optimal single-photon attack, turns out to be the lower bound for the Decoy-state BB84\,\cite{trushechkin2021_decoy}.

The second category includes protocols that encode information using pure coherent states and do not involve phase randomization. 
Protocols that fall under this category include Coherent One-Way protocols\,\cite{COWstucki2005fast, COWstucki2009high, COWkorzh2015provably,exp_COW}, early versions of the Differential Phase Shift protocols\,\cite{DPSinoue2003differential,DPStakesue2007quantum}, Strong Reference B92 protocols\,\cite{tamaki2008B92strongRef, tamaki2009strongRef, miroshnichenko2018strongRef} and Y-00 protocols\,\cite{Y00hirota2004quantum, barbosa2005information}.
The full security proof includes analysis of a general coherent attack, but~to upper-bound the key generation rate in these protocols, we may consider the beam-splitting (BS) \mbox{attack\,\cite{COWzero,kronberg2017analysis,kronberg2020quantum},} in~which Eve steals the portion of the signal expected to be lost in the line and then retransmits the remaining part to Bob through an ideal~channel.

In the following analysis, we deal with the influence of the loss control approach on two distinct protocols: the Decoy-State BB84 and Coherent One-Way protocols, each exemplary of one of the two identified groups.
The approach used in these protocols enables legitimate users to separate artificial losses from natural ones and, thus, precisely estimate the information available to Eve.
This refined estimation results in a less destructive reduction of the key length during the privacy amplification stage (compared to the original versions of the protocols).
We show that the improvement of privacy amplification in itself leads to a boost in the key generation rates for the considered protocols. 
Moreover, loss control allows legitimate users to utilize higher intensities of the signal states.
We analyze the influence of the intensities increase on the key generation rate and show that it leads to additional enhancement without sacrificing the everlasting security of the resulting key (see Section\,\ref{sec:beyond}). 

For the modified versions of the protocols, we primarily focus on the most feasible type of attack, the~local leakage modeled by the beam-splitter.
Given that photons are chargeless, the~only practicable method for an eavesdropper to interact with the transmitted signal is the altering the fiber medium.
Consequently, any attack strategies that deviate from local leakage attacks would necessitate significant changes to the line tomogram.
Such alterations would trigger the protocols to cease operations, assuming that the resolution of the line tomography is sufficient to enable users to detect and pinpoint any modifications to the fiber~medium.
  
\section{The BB84 protocol}\label{sec:BB84}
\begin{figure}[t]
\noindent\centering{
\includegraphics[width=0.99\columnwidth]{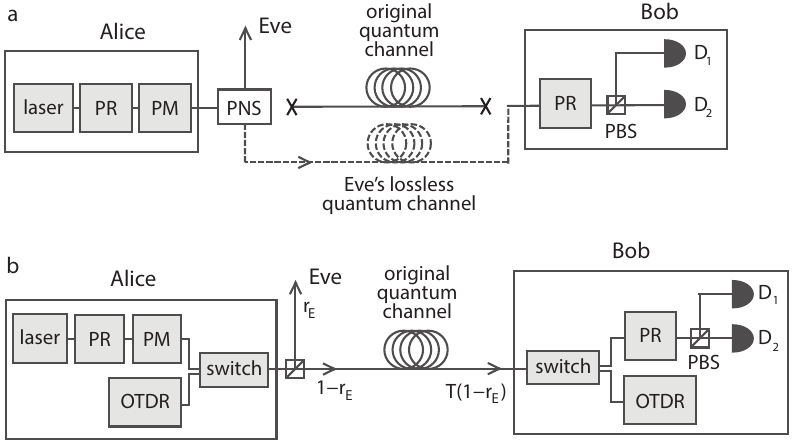}}
\caption{
\textbf{The BB84 protocol scheme.}
\textbf{a,} An original BB84 protocol.
Alice prepares a signal or a decoy state by utilizing the amplitude modulator\,(AM) and uses the polarization rotator\,(PR) to encode the information into the signal states.
Phase modulator\,(PM) randomizes the output states.
Bob chooses the basis by the PR and measures the arriving states by using the polarizing beam-splitter\,(PBS) and single-photon detectors $\text{D}_{1}$\,,$\text{D}_{2}$.
Eve performs the PNS\,(photon number-splitting) attack.
\textbf{b,} The enhanced BB84.
 Alice and Bob exploit the optical time-domain reflectometer (OTDR) to monitor losses in the line.
The switch element defines the working regime: generating a key or monitoring the line.
Eve introduces local intervention and intercepts the portion $r_\text{E}$ of the signal.}
\label{BB84sheme}
\end{figure}

We begin our analysis with the Decoy-State BB84, an~exemplary QKD protocol using phase randomization. 
Our objective is to determine the achievable key rates, first in the presence, and~then in the absence of the line tomography.
In this protocol, Alice encodes random bits utilizing four distinctive quantum states. 
These states constitute two sets of mutually unbiased orthonormal basises, namely the eigenbasis of the Pauli matrices, $\sigma_x$ and $\sigma_z$. 
The first set, X, consists of the $\ket{0}_x$ and $\ket{1}_x$ states; the second set, Z, contains $\ket{0}_z=\left(\ket{0}_x+\ket{1}_x\right)/\sqrt{2}$ and $\ket{1}_z=\left(\ket{0}_x-\ket{1}_x\right)/\sqrt{2}$ states.
Bob receives each state, guesses the basis with the 50\% success rate, and~measures the states accordingly. 
Once all measurements are completed, Alice discloses the correct basises, leading the users to discard any bit positions where the guessed and actual basises do not align. 
The remaining bit sequence undergoes post-processing to correct errors and eliminate the leaked information.
A visual layout of the protocol is depicted in Figure\,\ref{BB84sheme}a.

Initially, the~BB84 protocol was designed to be implemented via single-photon states\,\cite{BB84_orig}.
The security of such a realization was strictly proven\,\cite{Shor_preskill_security, Renner_thesis, Renner_article_security}, yet, practical implementation of a single-photon generator is the highly demanding engineering task and, in~practice, phase randomized weak coherent states are utilized as information carriers instead of single-photon pulses\,\cite{Experimental_Decoy_Free-Space_2007,kobayashi2014evaluation, boaron2018secure}.
Due to phase randomization, a~mixed state $\hat{\rho}$, a~statistical mixture of the Fock photon-number states, is sent each time instead of the pure coherent state $\ket{\gamma}$,
\begin{equation}
    \hat{\rho}
    =
    \frac{1}{2\pi}\int\limits_{0}^{2\pi}d\varphi \ket{\gamma e^{i\varphi}}\bra{\gamma e^{i\varphi}}
    =
    \sum\limits_{n=0}^{\infty}
    P_\gamma(n)\ket{n}\bra{n},
\end{equation}
where $\ket{n}$ is the $n$-photon Fock state and $P_{\gamma}(n)=e^{-|\gamma|^2} |\gamma|^{2n}/n!$ is the Poisson distribution\,\cite{Acin_Gisin_Scarani}.

Thus, with~the probability $P_\gamma(0)$, Alice sends a vacuum state to the optical line. With~the probability $P_\gamma(1)$, she sends a single-photon state and with the probability $1-P_\gamma(0)-P_\gamma(1)$, she sends a multi-photon state.
The presence of the multi-photon pulses allows Eve to perform the photon-number splitting (PNS) attack which involves replacing the original quantum channel with an ideal, lossless one and performing the non-demolition photon-number measurement\,\cite{PNS_attack_2000, PNS_Lutkenhaus, Acin_Gisin_Scarani}.
Eve collects one photon from each of the multi-photon pulses and stores the obtained photons in quantum memory until the basis reconciliation stage.
In order to compensate for the additional losses created during the PNS attack, Eve is to send the remaining photons to Bob via an ideal channel.
The condition $\braket{0_x|1_x}=\braket{0_z|1_z}=0$ enables Eve to distinguish between the logical bits `0' and `1' without a mistake by carrying out the measurement over kept photon in an appropriate basis. 
Hence, only single-photon pulses emitted by Alice's laser guarantee secure key distribution.
In order to estimate the contribution to the raw key provided by the single-photon pulses, legitimate users have to implement the decoy-state method\,\cite{Decoy1, Decoy2, trushechkin2021_decoy} which involves sending the additional pulses differing in intensities from the signal ones and analyzing their parameters at the receiver's~side.

\subsection{Secret key rate in modified BB84}
Adopting the proposed method based on the losses monitoring in the BB84 protocol, we build up the efficiency of the protocol, since Eve's attacks get restricted to inflicting local artificial losses that are not disruptive enough to be noticed by the line tomography.
An eavesdropper may store intercepted photons until basis reconciliation and apply optimal measurement to obtain full information about a bit. 
Thus, whenever %at least one
a photon is intercepted, Eve knows the bit value of the raw key: 
\begin{multline}
    I\left(\text{A,E}\right)
    =
    \sum\limits_{n=1}^{+\infty}P_{\sqrt{r_\text{E}}\gamma}(n)
    \!=\!
    1-P_{\sqrt{r_\text{E}}\gamma}(0)
    \\=
    1-e^{-r_\text{E}|\gamma|^2},
\end{multline}
where $r_\text{E}$ is the portion of the signal that Eve may seize imperceptibly.
In case of obtaining the conclusive result, receiving a non-zero number of photons, and~correctly guessing Alice's basis choice, Bob obtains full information about the bit value as well.
If $D_\text{AB}$ is the distance between Alice and Bob, the~transmittance of the whole optical line is determined as $T=10^{-\xi  D_\text{AB}}$, $\xi=0.02\,\text{km}^{-1}$.
Thus, the~probability of the conclusive result is 
\begin{multline}
    p_\checkmark=\frac{1}{2}\left(1-P_{\!\!\!\tiny{\sqrt{T(1-r_\text{E})}}\gamma}(0)\right)\\
    =\frac{1}{2}\left(1-e^{-T(1-r_\text{E})|\gamma|^2}\right).
\end{multline}
\vspace{-3pt}
\begin{figure}[t]
\noindent\centering{
\includegraphics[width=0.99\columnwidth]{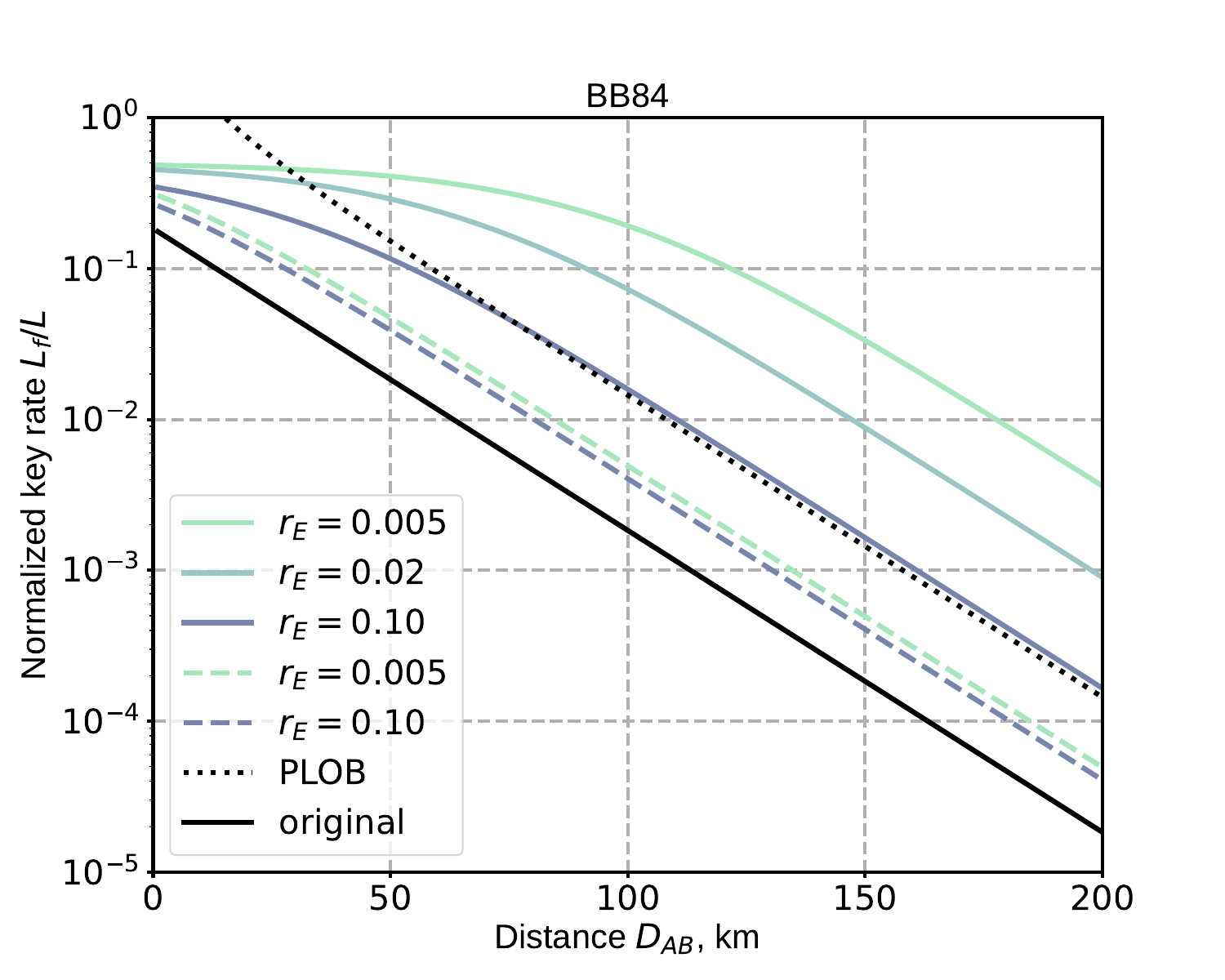}}
\caption{
\textbf{The BB84, secret key rate.}
The key generation rate as a function of the transmission distance $D_\text{AB}$ for enhanced Equation\,(\ref{BB84EnhKR}) and original Equation\,(\ref{BB84origKR}) versions of the BB84 protocol.
Different values of the leakage detection accuracy $r_\text{E}$ are considered: 0.005, 0.01, 0.10.
Solid lines stand for key rates of the enhanced version of the protocol for which the intensity maximizes Equation\,(\ref{BB84EnhKR}).
Optimal signal intensity $|\gamma|^2$ varies from 4 to 200 photons per pulse for the enhanced protocol (see Figure\,\ref{BB84number}).
Dashed lines stand for the key rates in the case when the loss control approach is applied, but~the intensity is not optimized and is taken the same as in the original version of the protocol.
The dotted line corresponds to the PLOB bound.
Here, the~errors in both bases are taken to be zero: $p^x_\text{err}=p^z_\text{err}=0$.}
\label{BB84rate}
\end{figure}
In  such a case, Bob's information about Alice's raw key, i.e.,~about the bit string Alice obtains after the post-selection stage, is $I(\text{A},\text{B})=1-(h_2(p^{x}_\text{err})-h_2(p^{z}_\text{err}))/2$, where $p^x_\text{err}$ and $p^z_\text{err}$ is error probabilities in the bases $x$ and $z$ respectively.
The analysis of errors' influence is provided in Appendix\,\ref{appendix_C}.
Applying the Devetak-Winter equation\,\cite{devetak2005distillation}, we explicitly calculate the final key generation rate $L_f$ for the modified version of the protocol
\begin{multline}
    L_f
    \geq
    L p_\checkmark\cdot\Big(I\left(\text{A},\text{B}\right)-I\left(\text{A},\text{E}\right)\Big)
    \\=
    \frac{L}{2}\cdot \Big(1-\frac{1}{2}h_2(p^x_\text{err})-\frac{1}{2}h_2(p^z_\text{err})\\
    -e^{-T(1-r_\text{E})|\gamma|^2}\Big)\cdot e^{-r_\text{E}|\gamma|^2},
    \label{BB84EnhKR}
\end{multline}
where $L$ is the rate at which Alice's random number generator produces bits.
This key generation rate estimation can be optimized over signal intensity $|\gamma|^2$.
Optimal intensity for Equation\,(\ref{BB84EnhKR}) is depicted in Figure\,\ref{BB84number} of Appendix \ref{appendix_A}.

\subsection{Comparison with Standard Decoy-State~BB84}
For the unmodified version of the Decoy-State BB84 protocol, we provide the upper bound on the secret key generation rate, see Appendix \ref{appendix_B} for details,
\begin{multline}
    L_f^\text{orig} \leq L\cdot \frac{1}{2} \Big[ T|\gamma|^2e^{-|\gamma|^2}\\
    -\left( 1-e^{-T|\gamma|^2}\right)\frac{1}{2}\left(h_2(p^x_\text{err})+h_2(p^z_\text{err})\right) \Big].
    \label{BB84origKR}
\end{multline}

Figure\,\ref{BB84rate} shows the dependence of the key rate on the distance between legitimate users $D_\text{AB}$ for original and modified versions of the BB84.
Precise estimation of Eve's information, which our method provides, enables legitimate users to exploit signal pulses with dozens of photons (see Figure\,\ref{BB84number}).
At a distance of 200\,km with today's reflectometers one may reach the detection accuracy of 0.5\% and enhance the performance of the protocol by about 100 times.
Even in the pessimistic case of the detection accuracy, $r_\text{E}=0.10$, one can boost the key generation rate several times.
Also, Alice and Bob may monitor the losses and not tune the average photon number in the signal pulses.
In this case, legitimate users, modifying only privacy amplification (dashed lines at Figure\,\ref{BB84rate}), increase the key generation rate more than 2 times.
In comparison, the~asymptotic behavior of the key rate provided by the PLOB at a 200 km distance is limited to values around $10^{-4}$, which is the order of magnitude lower than the rates achievable with the appropriate level of leakage detection accuracy.
To conclude, a~significant boost can be achieved without modifying the QKD equipment, one only needs to monitor the losses in the line and carry out more rational privacy amplification.
It makes the QKD protocols available for a wide range of users at the present~time.

 \section{Coherent One-Way protocol}\label{sec:COW}
\begin{figure}[t]
\noindent\centering{
\includegraphics[width=0.99\columnwidth]{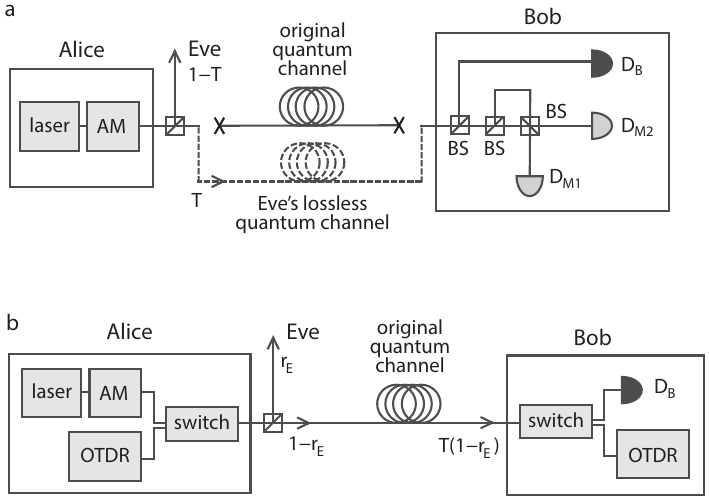}}
\caption{
\textbf{COW protocol scheme.}
(\textbf{a}) Original COW.
Alice prepares coherent states using the laser and adjusts the signal’s amplitude with the amplitude modulator\,(AM).
BS stands for a beam-splitter.
The detector $D_\text{B}$ monitors the arrival time of the signals.
Detectors $\text{D}_\text{M2}$ and $\text{D}_\text{M1}$ check weather the arriving sequence has the from $\ket{\gamma}\ket{\gamma}$.
Eve performs the BS attack.
(\textbf{b}) Enhanced COW.
Alice and Bob exploit OTDR\,(optical time-domain reflectometer) to monitor losses in the line.
The switch element defines the working regime: generating a key or monitoring the line.
Eve introduces local intervention and intercepts the portion $r_\text{E}$ of the signal.
}
\label{COWsheme}
\end{figure}

In the COW protocol, see Figure\,\ref{COWsheme}a, Alice utilizes an attenuated laser and prepares coherent states with the intensity $|\gamma|^2$ to encode a random bit string into two-pulse sequences composed of the non-empty and empty pulses, $0 \rightarrow \ket{0}\ket{\gamma},\, 1 \rightarrow \ket{\gamma}\ket{0}$.
Through the optical fiber, prepared states are sent to Bob, who measures them by single photon detector $\text{D}_\text{B}$.
The detector monitors the pulses' arrival time, according to which Bob makes bit decisions.
Since the coherent and vacuum states are non-orthogonal, the~detector $\text{D}_\text{B}$ sometimes does not click on the non-empty pulses. Bob considers such measurement results as inconclusive and discards them at the post-selection~stage.

Bob's scheme also includes the interferometer, the~long arm of which has the length assuring that the two non-empty neighboring pulses interfere at the last beam-splitter (see Figure\,\ref{COWsheme}a).
Thus, the~detector $\text{D}_\text{M2}$ does not react to the arriving pulse sequence of the form $\ket{\gamma}\ket{\gamma}$, contained in the sequence corresponding to logical bits ``01'' ($\ket{0}\ket{\gamma}\ket{\gamma}\ket{0}$).
If an eavesdropper blocks a part of such a sequence, the~visibility between detectors $\text{D}_\text{M2}$ and $\text{D}_\text{M1}$ will inevitably change.
As a result, the~attacks that include blocking a part of transmitted pulses in the original scenario of the COW can be potentially detected by sending the control states $\ket{\gamma}\ket{\gamma}$.

\subsection{Secret key rate in modified COW}
Next, we delve into the analysis of the COW in the context of the loss control approach that restricts the eavesdropper's actions to local interventions in the line (see Figure\,\ref{COWsheme}b).
If $r_\text{E}$ is the minimal detectable artificial leakage, an~eavesdropper has to measure the states from the ensemble $\ket{\sqrt{r_\text{E}}\gamma}\!\ket{0},\,\ket{0}\!\ket{\sqrt{r_\text{E}}\gamma}$, the~Holevo quantity\,\cite{Holevo} $\chi$ of which upper-bounds the mutual information between Alice and Eve
\begin{equation}
    I(\text{A,\,E})\leq\chi= h_2\left( \frac{1-| \braket{0|\sqrt{r_\text{E}}\gamma} |^2}{2} \right).
\end{equation}

In this context, Eve does not introduce any errors in the raw key.
Hence, when Bob obtains a conclusive measurement result, the~mutual information between Alice and Bob is one bit: $I(\text{A,\,B})=1$, here we neglect the dark counts in detectors and other equipment’s imperfections.
The probability that measures a single bit-carrying signal through which Bob gets a conclusive outcome is determined by the Poisson statistics of the coherent state $\ket{\sqrt{T(1-r_\text{E})}\gamma}$
\begin{equation}
    p_\checkmark = 1 - \left| \braket{0|\sqrt{T(1-r_\text{E})}\gamma} \right|^2 = 1-e^{ -T(1-r_\text{E})|\gamma|^2 }.
\end{equation}

After post-selection and privacy amplification procedures, the~length of the final key $L_f$ is also calculated according to the Devetak-Winter approach
\begin{multline}
    L_f = p_\checkmark L \cdot \left( I(\text{A,\,B})-I(\text{A,\,E}) \right) \\
    \geq p_\checkmark L \cdot \left( 1-h_2(p_\text{err})-h_2\left( \frac{1-e^{-r_\text{E}|\gamma|^2}}{2} \right) \right).
    \label{COWrateEqMod}
\end{multline}
\subsection{Comparison with original COW}
To find an upper bound of the key rate in the original protocol, it is sufficient to consider any of the possible eavesdropping attacks.
We consider the beam-splitting (BS) attack which is not the most optimal eavesdropping on the COW but which is a suitable reference since in our approach such an attack appears to be a basic one.
As an example of a more powerful attack, we mention the one based on the soft filtering operation\,\cite{kronberg2020quantum}. 
We analyze the BS attack, in~which Eve replaces the optical line with the ideal one and intercepts the signal's part expected to be lost.
Thus, Eve, simulating the natural losses in the channel with a proportion of $1-T$, has to distinguish between states $\ket{\sqrt{1-T}\gamma}\ket{0},\ket{0}\ket{\sqrt{1-T}\gamma}$.
The maximum information that Eve can obtain, on~average, about Alice’s bit is bounded by the Holevo quantity, which for pure equiprobable states has the form
\begin{equation}
    I(\text{A,\,E})\leq\chi= h_2\left( \frac{1-| \braket{0|\sqrt{1-T}\gamma} |^2}{2} \right).
\end{equation}

Similarly to the previous consideration of the local leakage $r_\text{E}$, errors will not occur for the conclusive results at Bob's side ($I(\text{A,\,B})=1$).
The probability of a conclusive outcome is determined by the Poisson statistics of the coherent state $\ket{\sqrt{T}\gamma}$
\begin{equation}
    p_\checkmark = 1- \left| \braket{0|\sqrt{T}\gamma} \right|^2 = 1-e^{ -T|\gamma|^2 }.
\end{equation}
Thus, the~resulting key generation rate for the BS-attack can be estimated as follows
\begin{multline}
    L_f^\text{orig} \leq p_\checkmark L \cdot \left( I(\text{A,\,B})-I(\text{A,\,E}) \right)\\
    \!\!\!= p_\checkmark L \left( 1-h_2\left(p_\text{err}\right)- h_2 \left( \frac{1-e^{-(1-T)|\gamma|^2}}{2} \right) \right).
    \label{origCOWkr}
\end{multline}

\begin{figure}[t]
\noindent\centering{
\includegraphics[width=0.99\columnwidth]{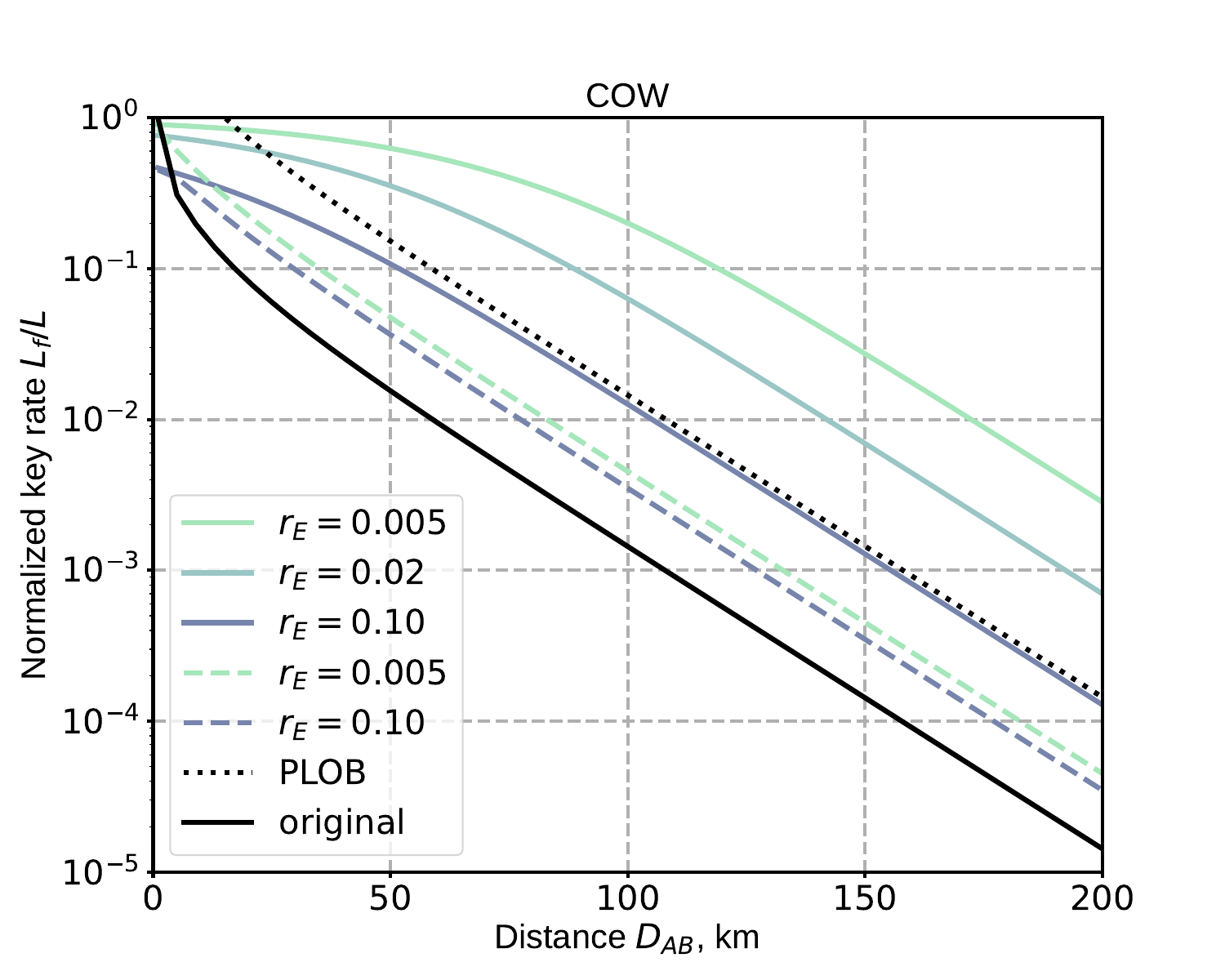}}
\caption{
\textbf{COW, secret key rate.}
The key generation rate as a function of the transmission distance $D_\text{AB}$ for enhanced Equation\,(\ref{COWrateEqMod}) and original Equation\,(\ref{origCOWkr}) versions of the COW protocol.
Different values of the leakage detection accuracy $r_\text{E}$ are considered: 0.005, 0.01, 0.10.
Solid lines stand for key rates of the enhanced version of the protocol for which the intensity maximizes Equation\,(\ref{COWrateEqMod}).
Optimal signal intensity $|\gamma|^2$ varies from 2 to 100 photons per pulse for the enhanced protocol (see Figure\,\ref{COWnumber}).
Dashed lines stand for the key rates in the case when the loss control approach is applied, but~the intensity is not optimized and is taken the same as in the original version of the protocol.
The dotted line corresponds to the PLOB bound.
Here, we consider zero-error case $p_\text{err}=0$.}
\label{COWrate}
\end{figure}
\unskip

Figure\,\ref{COWrate} displays the result of numerical simulations for the COW with the loss control approach and the original version of the protocol.
The normalized key rate $L_f/L$ is depicted as a function of the transmission distance $D_\text{AB}$ for different values $r_\text{E}$: 0.005, 0.01, 0.10.
For each portion of the stolen signal, $r_\text{E}$, and~distance $D_\text{AB}$, the~optimal intensity $|\gamma|^2$ is found to maximize the key rate determined by Equation\,(\ref{COWrateEqMod}).
Applying the loss control approach allows for an increase in the intensity of signal pulses up to dozens of photons, see Figure\,\ref{COWnumber} of Appendix \ref{appendix_A}, and~significantly boosts the key rate compared to the original version of the COW.
The modified protocol produces a higher key rate even in the pessimistic case where Eve gets stolen about 10\% of the signal.
At a distance of 200\,km and leakage detection precision $r_\text{E}=0.005$, the~COW performance can be improved about 100 times in terms of key rates.
Without modifying the signal average photon number, legitimate users can double the achievable key generation rate, see dashed lines at Figure\,\ref{COWrate}.
Again, if~we are able to ensure the loss control of the level $r_\text{E}<0.1$, we can overcome the PLOB~bound.

 \section{Loss Control Compared to the Decoy-State~Method}\label{sec:comparison}
The decoy-state method, utilized in the BB84, allows for legitimate users to detect, by~estimating the number of non-blocked single-photon pulses, Eve's attacks using the complete blocking of some transmitted states.
This method, employing decoy pulses with intensities different from ones in the bit-encoding states, is mostly aimed to cope with the PNS attack that appears to be highly relevant in the context of the experimental QKD realizations.
Our approach also exploits special test pulses in addition to the signal ones but allows us to detect a wider spectrum of eavesdropping attacks.
Based on the natural losses analysis and on the infeasibility of their exploitation, we step beyond the conventional decoy-state method and acquire the ability to determine the portion of the signal available to Eve.
Precise estimate of the eavesdropper's information in the proposed approach leads to the less destructive compression of the key during the privacy amplification stage and allows for the utilization of bit-encoding states with hundreds of~photons.

 \section{Beyond the Prepare-and-Measure~QKD}\label{sec:beyond}
In our work, we concentrate on the enhancement of the prepare-and-measure QKD protocols.
Notably, protocols belonging to other classes can be modified as well.
One can consider the Twin-Field (TF) QKD protocol\,\cite{TF} as an example.
The working principle of the TF-QKD lies in sending the quantum state from Alice and Bob to the intermediate measurement point.
The setting is equivalent to using one quantum repeater that results in overcoming the PLOB bound. 
Applying our approach to the TF-QKD protocol, legitimate users may monitor the losses from both ends of the line and, thus, they can produce secret keys with rates significantly exceeding the ones achieved in the state-of-the-art realizations\,\cite{TFexperiment2021,TFexperiment2022,TFexperiment2023,tf1,tf2,tf4,tf5,tf6}.

It is important to clarify that our approach is formulated in the framework of the device-dependent QKD (according to the definition by R.\,Renner~\cite{Renner_Wolf}), protocols that rely ``on the exact specification of the deployed devices for their security proof.'' 
Examples of such protocols are all the prepare-and-measure and the entanglement-based QKD. 
Our method is more device-dependent than conventional QKD protocols since we also rely on the OTDR and transmittometry devices (see Table\,\ref{table_compare}). 
Thus, protocols that belong to the device independent QKD~\cite{mdi1,mdi2,mdi3,mdi4,mdi5} including the MDI variant of the Twin-Field QKD\,\cite{tf3} are beyond the framework of our work.
Nevertheless, the~key generated using our approach is everlastingly secure, meaning that the key remains secret in time even against the non-existing technologies or attacks\,\cite{portmann2022security, unruh2013everlasting,stebila2010case,alleaume2010quantum}.

\setlength{\tabcolsep}{0.5em} % for the horizontal padding
{\renewcommand{\arraystretch}{1.2}% for the vertical padding
\begin{table}[t]
\centering{
\begin{tabular}{lcccc}
\hline && \begin{tabular}[c]{@{}c@{}}Secret\\key rate\end{tabular}   & \begin{tabular}[c]{@{}c@{}}Everlasting\\ security\end{tabular} & \begin{tabular}[c]{@{}c@{}}Device \\ dependency\end{tabular} \\ \hline
\begin{tabular}[c]{@{}l@{}}Post-quantum\\ cryptography\end{tabular}            && high            & \xmark & --- \\ \hline
\begin{tabular}[c]{@{}c@{}}Point-to-point \\ QKD\end{tabular} &&  \begin{tabular}[c]{@{}c@{}}relatively \\ low\end{tabular}  & \checkmark & \begin{tabular}[c]{@{}c@{}}relatively \\ high\end{tabular}                                              \\ \hline
MDI QKD && low & \checkmark & medium \\ \hline
\begin{tabular}[c]{@{}l@{}}Point-to-point \\QKD with \\loss control\end{tabular} && \begin{tabular}[c]{@{}c@{}}relatively \\ high\end{tabular} & \checkmark & high \\ \hline
\end{tabular}
\caption{{\textbf{Cryptography approaches comparison.}
Post-quantum cryptography, Point-to-point QKD, MDI QKD and Point-to-Point with loss control have different properties with regard to key rate, everlasting security and device dependency.
In the table, ``\checkmark'' means that key generated by a particular method is secure in time and resistant to both software and hardware future developments. }}
}
\label{table_compare}
\end{table}
}

\section{Discussion}\label{sec:discussion}
The fundamental PLOB bound\,\cite{PLOB} of the key rate arises from the fact that all losses occurring in a quantum channel can be effectively measured by an eavesdropper.
In this paradigm, protocols that do not exploit quantum repeaters provide relatively small key rates at hundreds of kilometers.
Aimed at deflecting known attacks associated with channel losses, our approach acknowledges that it is unfeasible to gain information from natural losses.
At the same time, we assert that any intrusive actions by a global eavesdropper, like substituting a channel with an ideal one, can be detected via line tomography.
As a result, Alice and Bob separate the natural losses from local artificial leakages and determine the part of the signal available to an eavesdropper.
Legitimate users, thus, may carry out privacy amplification with a less destructive reduction of the key length.
This complements the results of~\cite{MDPI_Entropy_Improving_the_perfomance} devoted to secret key rate increase observed under another kind of technological constraints imposed on a potential eavesdropper. While, in~this paper, we do not consider the efficiency of the methods combination, we see it as a promising area for future~research.

Intriguingly, it turns out that our approach also allows the users to increase the average photon number in bit-encoding states.
This method, as~we illustrate in the context of the COW and BB84 protocols, leads to a substantial enhancement in the key generation rate.
At a distance of 200\,km our method provides a 100 times higher key rate than the original versions of the protocols.
Based on the proposed approach, with~the appropriate level of leakage detection, we can potentially overcome the PLOB bound without exploiting quantum repeaters or devices acting as repeaters, for~example, amplifiers.
To overcome PLOB, we do not introduce mathematical modifications in existing analysis but change the model of the key generation process by relying on physically motivated~assumptions.

Notably, while our method is applied to the existing QKD recipes and is based on the physics-motivated assumptions about the eavesdropper's opportunities, we leave an information-theoretical security proof for the increased photon number method, as~well as a more formal security treatment, taking into account non-asymptotic effects, for~our forthcoming~work.

\vspace{6pt} 

\begin{appendices}
  
 \section{Optimal intensities for modified versions of BB84 and COW}\label{appendix_A}
\begin{figure}[t]
\noindent\centering{
\includegraphics[width=0.99\columnwidth]{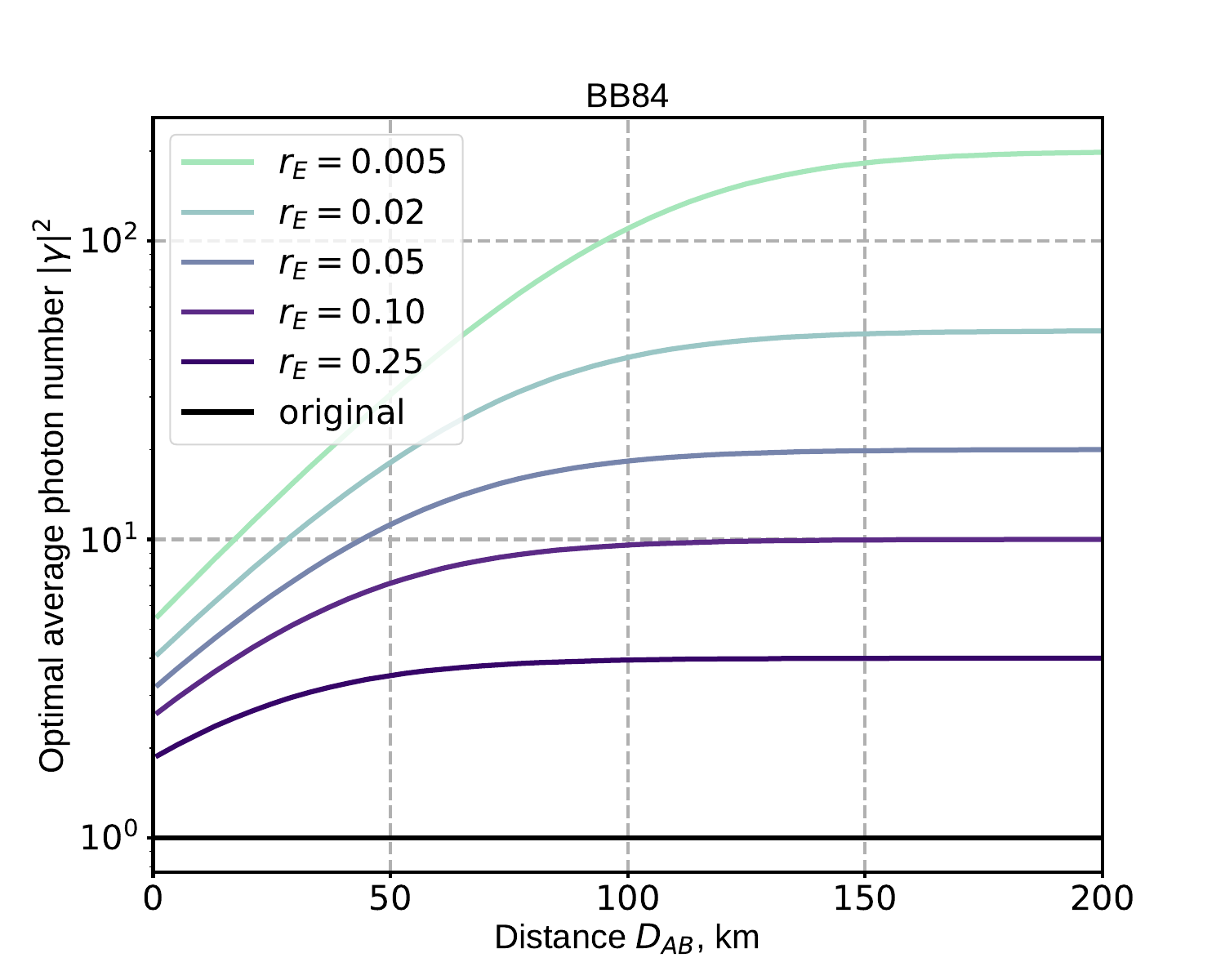}}
\caption{
\textbf{Optimal photon number, BB84.}
The average photon number $|\gamma|^2$ which maximizes the key generation rate for both, original and enhanced, versions of the BB84 is a function of the transmission distance $D_\text{AB}$.
For the enhanced version, $|\gamma|^2$ maximizes Eq.\,(\ref{BB84EnhKR}) for different values of leakage $r_\text{E}$: 0.005, 0.01, 0.10 (color lines).
For the original version, $|\gamma|^2$ that maximizes Eq.\,(\ref{BB84origKR}) is equal to 1 (black line).
}
\label{BB84number}
\end{figure}

\begin{figure}[t]
\noindent\centering{
\includegraphics[width=0.99\columnwidth]{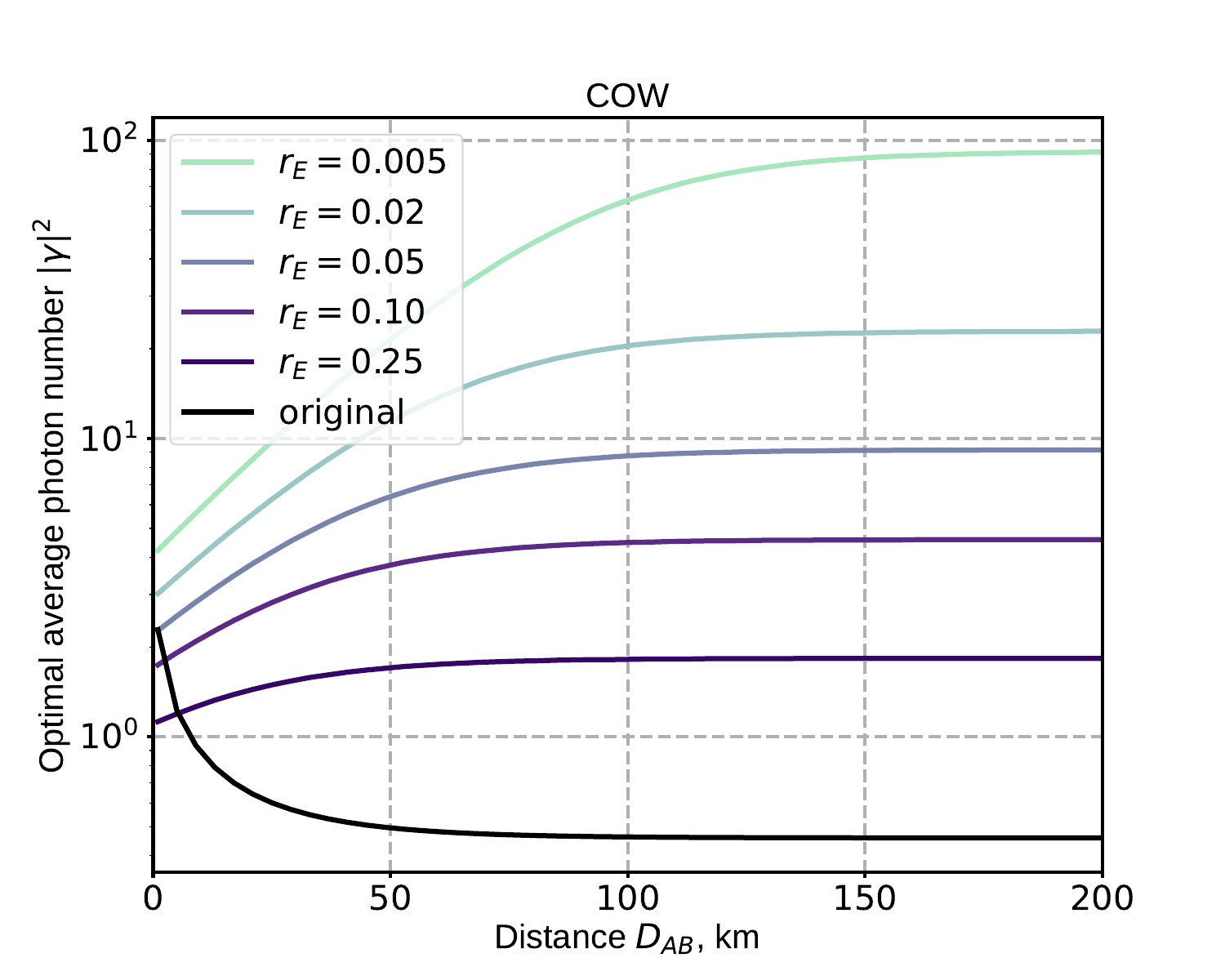}}
\caption{
\textbf{Optimal photon number, COW.}
The average photon number $|\gamma|^2$ which maximizes the key generation rate for original and enhanced versions of COW is a function of the transmission distance $D_\text{AB}$.
For the enhanced version $|\gamma|^2$ maximizes Eq.\,(\ref{COWrateEqMod}) for different values of leakage $r_\text{E}$: 0.005, 0.01, 0.10 (color lines).
For the original version $|\gamma|^2$ maximizes Eq.\,(\ref{origCOWkr})(black line).}
\label{COWnumber}
\end{figure}

Each QKD setup parameter should maximize the resulting key generation rate.
In the original versions of the protocols, the~optimal average photon number $|\gamma|^2$ is about \mbox{1 photon,} which dramatically suffers from channel decay at hundreds of kilometers. 
Taking our approach, precise estimation of Eve's information allows legitimate users to utilize bit-encoding quantum states with much higher average photon numbers.
Figure\,\ref{BB84number} shows $|\gamma|^2$ that maximizes key rate Equation\,(\ref{BB84EnhKR}) for the enhanced version of BB84.
It, of~course, depends on the leakage $r_\text{E}$ and varies from several photons to more than 100 photons per pulse.
The optimal intensity for COW is represented in Figure\,\ref{COWnumber}.

\unskip
  
 \section{Appendix B. Upper bound on key rate in Decoy-State BB84}\label{appendix_B}
Due to the threat of the PNS attack, multi-photon pulses are insecure since Eve may keep one photon in quantum memory and re-send the rest to Bob through an ideal channel.
Secure key generation is guaranteed only by one-photon pulses, the portion of which Eve can block.
To detect such eavesdropping actions and estimate the number of one-photon pulses that reached Bob, Hwang introduced the decoy-state method\,\cite{hwang2003quantum}.
In the decoy-state paradigm, Alice sends special pulses with different intensities compared to signal ones.
The analytical expression\,\cite{Decoy1,Decoy2} for the maximum length of the secret key that can be achieved is
\begin{equation}
    L_f^\text{orig}\!=\!L\frac{1}{2}\left[Q_1(1 - h_2(e_1))-Q f(p_\text{err}) h_2(p_\text{err})\right],
    \label{decoy_secret_key_rate}
\end{equation}
where $Q$ is the gain of signal states (the probability that a signal state will be detected by Bob) and $p_\text{err}$ is the quantum bit error rate (QBER); $f(p_\text{err})\geq 1$ is the efficiency of an error-correction procedure with the Shannon limit $f(p_\text{err})=1$.
The quantity $Q_1$ is the gain of single-photon states (a joint probability that a single-photon pulse was emitted by Alice and detected by Bob), and $e_1$ is the error rate on single-photon pulses. 
The quantities $Q$ and $p_\text{err}$ can be measured in the experiment, while $Q_1$ and $e_1$ cannot be observed directly due to the fact that Bob is not able to distinguish between photons that originated from the single-photon and multi-photon pulses.
The values $Q_1$ and $e_1$ can be estimated by analyzing the parameters of decoy pulses on Bob's side\,\cite{Decoy1, Decoy2}.

We consider a situation when Eve is basically absent and does not conduct any attack at all, but the legitimate users do not know that and have to estimate the key generation rate fairly to be on the safe side.
This approach enables us to estimate the upper bound on the key rate ensuring its independence of the experimentally observed parameters.
To find an upper bound on Eq.\,(\ref{decoy_secret_key_rate}), one can use non-negativity of binary entropy $h_2$: $h_2(e_1)\geq 0$ and $h_2(p_\text{err})\geq 0$, and get that $L_f^\text{orig}\leq LQ_1/2$.
Eve's activity causes the decrease of the gain of single-photon states $Q_1$, thus, it is maximum in Eve's absence: $Q_1\leq T\cdot |\gamma|^2e^{-|\gamma|^2}$. 
Consequently, the upper bound for the length of a shared secret is
\begin{equation}
    L_f^\text{orig}\leq L\cdot\frac{1}{2}T\cdot |\gamma|^2e^{-|\gamma|^2}.
\end{equation}
The analysis of this expression shows that its maximum is attained when $|\gamma|=1$.
When we are interested in the key rate dependence on the observed error, we should not discard the second term in Eq.\,(\ref{BB84origKR}).
The gain of signal states is determined by the average photon number in bit-encoding states $|\gamma|^2$ and transmittance $T$ and equals to $Q=1-e^{-T|\gamma|^2}$.
For the upper bound we consider $e_1=0$.
Then, key rate as a function of the error probability takes the form Eq.(\ref{BB84origKR}).

 \section{Appendix C. Error analysis in BB84}\label{appendix_C}
\begin{figure}[t]
\noindent\centering{
\includegraphics[width=0.99\columnwidth]{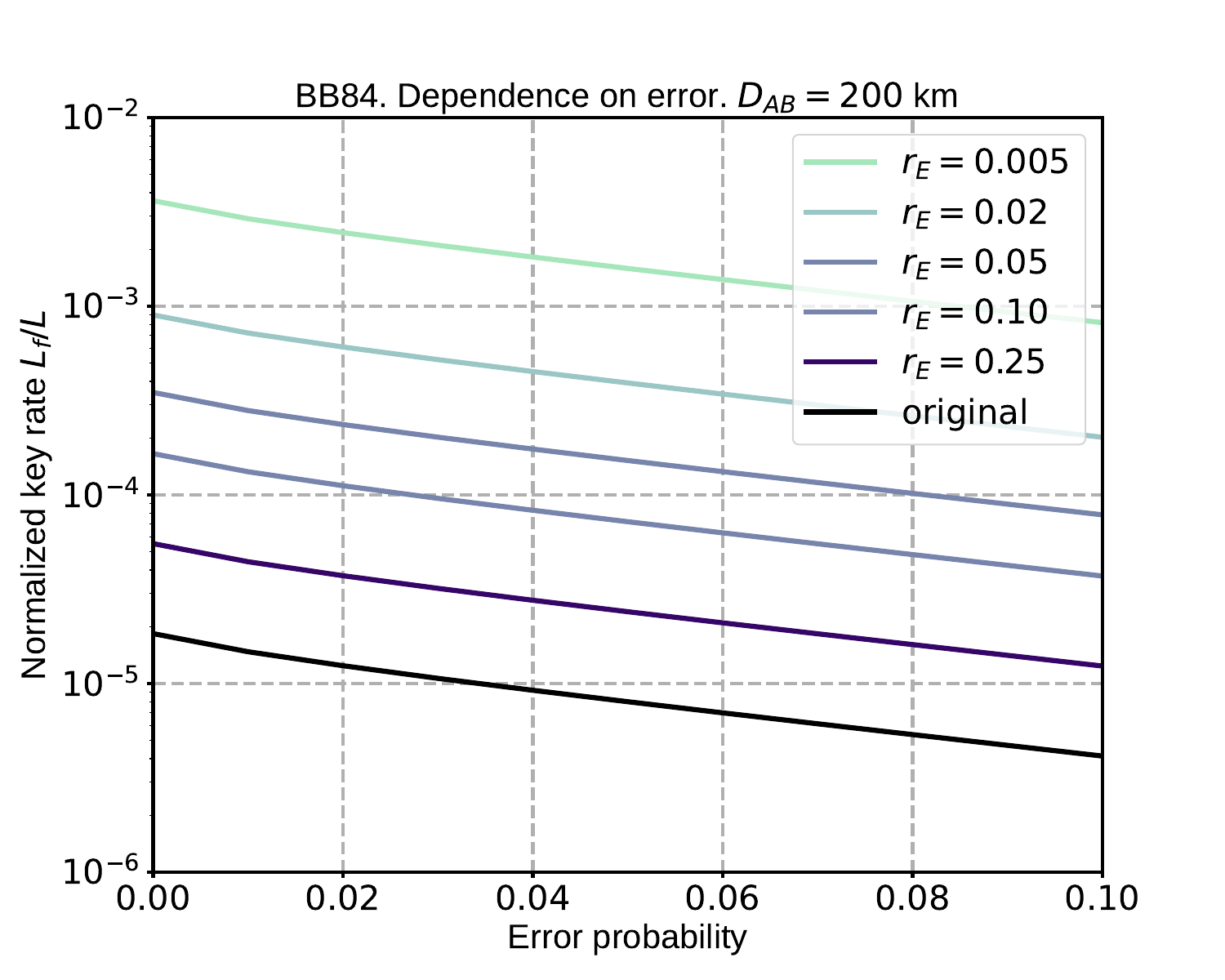}}
\caption{
\textbf{Dependence on error probability.}
The key generation rate as a function of error probability for both, original and enhanced, versions of BB84.
The transmission distance is $D_\text{AB}=200$\,km.
For the enhanced version, considered values of leakage $r_\text{E}$: 0.005, 0.01, 0.02, 0.10, 0.25 (color lines).
For the original version, key rate is calculated according to Eq.\,(\ref{BB84origKR}) (black line).
Error probability is expected to be the same in both bases.}
\label{BB84errorProb}
\end{figure}

In the modified version of BB84, errors may appear due to the imperfections of the line and other quantum devices rather than Eve's actions.
The dependence of the key generation rate on the errors for both versions of the protocol is represented at Fig.\,\ref{BB84errorProb}.
For the accuracy of leakage detection $r_\text{E}=0.005$ and error probability of 10\% the modified version of the protocol provides 200 times higher key generation rate as in the analysis without errors.

\end{appendices}

\newpage
%\bibliography{re.bib}

\end{document}